\documentclass[a4paper]{article}

\usepackage{INTERSPEECH2019}

\usepackage{graphicx}
\usepackage{amssymb,amsmath,bm}
\usepackage{textcomp}

\usepackage{epstopdf}
\usepackage{tabularx}
\usepackage{multirow}
\usepackage{textcomp}

\usepackage{lipsum} 
\usepackage{enumitem}
\setlist{nosep}

\graphicspath{{./image/}}

\usepackage{color}

\usepackage{bm}
\usepackage{calc}
\usepackage{balance}

\usepackage{enumitem}

\def\vec#1{\ensuremath{\bm{{#1}}}}

\title{Spatio-Temporal Attention Pooling for Audio Scene Classification}
\name{Huy Phan$^{*1}$, Oliver Y. Ch\'en$^2$, Lam Pham$^1$, Philipp Koch$^3$, Maarten De Vos$^2$, \\Ian McLoughlin$^1$, Alfred Mertins$^3$}
\address{
  $^1$School of Computing, University of Kent, UK\\
  $^2$Department of Engineering Science, University of Oxford, UK\\
  $^3$Institute for Signal Processing, University of L\"ubeck, Germany}
 \email{\textnormal{$^{*}$Corresponding email:} h.phan@kent.ac.uk}

\begin{document}

\maketitle
\begin{abstract}
Acoustic scenes are rich and redundant in their content. In this work, we present a spatio-temporal attention pooling layer coupled with a convolutional recurrent neural network to learn from patterns that are discriminative while suppressing those that are irrelevant for acoustic scene classification. The convolutional layers in this network learn invariant features from time-frequency input. The bidirectional recurrent layers are then able to encode the temporal dynamics of the resulting convolutional features. Afterwards, a two-dimensional attention mask is formed via the outer product of the spatial and temporal attention vectors learned from two designated attention layers to weigh and pool the recurrent output into a final feature vector for classification. The network is trained with \emph{between-class} examples generated from between-class data augmentation. Experiments demonstrate that the proposed method not only outperforms a strong convolutional neural network baseline but also sets new state-of-the-art performance on the LITIS Rouen dataset.

\end{abstract}

\section{Introduction}

Audio scene classification (ASC) is one of the main tasks in environmental sound analysis. It allows a machine to recognize a surrounding environment based on its acoustic sounds \cite{Mesaros2018}. One way to look at an audio scene is to consider its foreground events mixed with its background noise. Due to the complex content of audio scenes, it is challenging to classify them correctly, as classification models tend to overfit the training data. A good practice in audio scene classification is to split a long recording (e.g. 30 seconds) into short segments (a few seconds long) \cite{phan2017c, Mun2017a, Han2017, Han2016}. By this, we increase the data variation and a classification model can be trained more efficiently with a large set of small segments rather than a small set of the whole long recordings. Classification of long recordings is then achieved by aggregating classification results across the decomposed short segments. 

Similar to many other research fields, using deep learning for the ASC task has become a norm. Convolutional neural network (CNNs) \cite{phan2017b, Valenti2016, phan2017a, Han2016, Han2017, Yin2018, Zhang2018, Nguyen2018} are most commonly used deep learning techniques, thanks to their feature learning capability. Leveraging the nature of audio signals, sequential modelling with recurrent neural networks (RNNs) \cite{phan2017c, Vu2016, Guo2017} and temporal transformer networks \cite{Zhang2018} have also demonstrated results on par with the convolutional counterparts. The deep learning models were trained either on time-frequency representations, such as log Mel-scale spectrograms \cite{Han2017, Nguyen2018}, 
or high-level features like label tree embedding features \cite{phan2017b, phan2017c}. Mitigation of overfitting effects via feature fusion \cite{Han2017, phan2017b, phan2017c} and model fusion \cite{Nguyen2018, Phan2019c, Mesaros2018, Yin2018} has also been extensively explored.

Given the rich content of acoustic scenes, they typically contain a lot of irrelevant and redundant information. This fact naturally gives rise to the question of how to encourage a deep learning model to automatically discover and focus on discriminative patterns and suppress irrelevant ones from the acoustic scenes for better classification. We seek to address that question in this work using an attention mechanism \cite{Luong2015b}. To this end, we propose a spatio-temporal attention pooling layer in combination with a convolutional recurrent neural network (CRNN), inspired by their success in the audio event detection task \cite{Cakir2017,Phan2019b}. The convolutional layers of the CRNN network are used to learn invariant features from time-frequency input, whose temporal dynamics are subsequently modelled by the upper bidirectional recurrent layers. Temporal soft attention \cite{Luong2015b} has usually been coupled with a recurrent layer to learn a weighting vector to combine its recurrent output vectors at different time steps into a single feature vector. However, spatial attention (i.e. attention on the feature dimension), and hence, joint spatio-temporal attention, have been left uncharted. With the proposed spatio-temporal attention layer, we aim to learn a two-dimensional attention mask for spatio-temporal pooling purpose. The rationale is that those entries of the recurrent output that are more informative should be assigned with strong weights and vice versa. We expect discriminative features to be accentuated in the induced feature vector, while irrelevant ones to be suppressed and blocked after the spatio-temporal attention pooling. In addition, we harness between-class data augmentation \cite{Tokozume2018} to generate between-class examples to better train the network to minimize the Kullback-Leibler (KL) divergence loss. 

\section{The proposed CRNN with spatio-temporal attention pooling}

\subsection{Input}
\label{ssec:input}
Following the common practice in ASC \cite{phan2017c,Mun2017a, Han2017, Han2016}, we decompose an audio snippet, which is 30 seconds long in the experimental LITIS Rouen dataset \cite{Rakotomamonjy2015} (cf. Section \ref{ssec:dataset}), into non-overlapping 2-second segments. It has been shown in previous works that an auxiliary channel which is created by excluding background noise from an audio recording is also helpful for the classification task, as the prominent foreground events of the scene are exposed more clearly to a network \cite{Han2017,phan2017c,phan2017b}. Hence, we create such an auxiliary channel by subtracting background noise using the minimum statistics estimation and subtraction method \cite{Martin2001}.

A 2-channel short audio segment is then transformed into the time-frequency domain, e.g. using log Mel spectrogram, to obtain a multi-channel image $\mathbf{S}\in \mathbb{R}^{M\times T \times K}$, where $M$, $T$, and $K=2$ denote the number of frequency bins, the number of time indices, and the number of channels, respectively (cf. Section \ref{ssec:features} for further detail).

\begin{figure} [!t]
	\centering
	\includegraphics[width=1\linewidth]{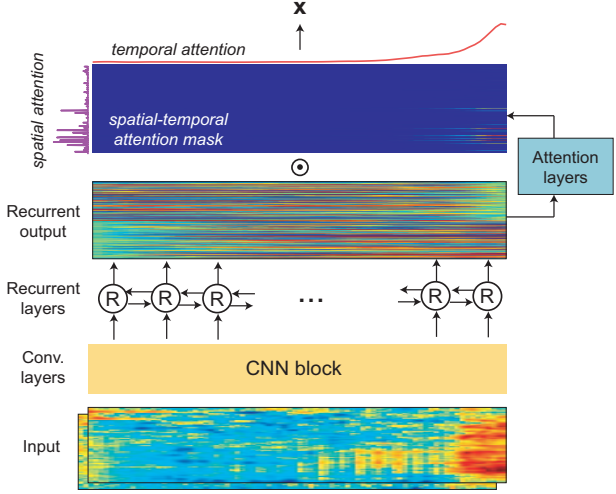}
	\caption{Overview of the proposed CRNN with spatio-temporal attention pooling.}
	\label{fig:overview}
\end{figure}

\subsection{Convolutional layers}
\label{ssec:convolution}
The convolutional block of the network consists of three convolutional layers followed by three max-pooling layers. For clarity, we show the configuration of the convolutional and max-pooling layers in Table \ref{tab:layers} alongside their corresponding resulting feature maps.

For each convolutional layer, after the convolution operation, Rectified Linear Unit (ReLU) activation \cite{Nair2010} and batch normalization \cite{Ioffe2015} are exercised on the feature map. The number of convolutional filters of a convolutional layer is designed to be the double of its preceding layer, i.e. $64 \rightarrow 128 \rightarrow 256$, in order to gain the representation power when the spectral size gets smaller and smaller after the pooling layers. Note that zero-padding (also known as \emph{SAME} padding) is used during convolution to keep the temporal size unchanged (i.e. always equal to $T$).

With the pooling kernel size $4 \times 1$ and a stride $1 \times 1$, the max pooling layers are only effective on the frequency dimension to gain spectral invariance. As a consequence, the spectral dimension is reduced from $M$ of the original input to $\frac{M}{4} \rightarrow \frac{M}{16} \rightarrow \frac{M}{64}$ after the three pooling layers, respectively. The last resulting feature map of size $\frac{M}{64} \times T \times 256$ is reshaped to $\mathbf{O} \in \mathbb{R}^{F \times T}$, where $F = \frac{M}{64}\times256$, to present to the upper recurrent layers of the network which will be elaborated in the following section.

\subsection{Bidirectional recurrent layers with spatio-temporal attention pooling}
\label{ssec:recurrent}

In the context of sequential modelling with recurrent layers, the convolutional output $\mathbf{O} \in \mathbb{R}^{F \times T}$ is interpreted as a sequence of $T$ feature vectors $(\mathbf{o}_1, \mathbf{o}_2, \ldots, \mathbf{o}_T)$ where each $\mathbf{o}_t \in \mathbb{R}^F$, $1 \le t \le T$. A bidirectional recurrent layer then reads the sequence of convolutional feature vectors into a sequence of recurrent output vectors $\mathbf{Z} \equiv (\mathbf{z}_1, \mathbf{z}_2, \ldots, \mathbf{z}_T)$, where
\begin{align}
\mathbf{z}_t &= [\mathbf{h}^{\text{b}}_t \oplus \mathbf{h}^{\text{f}}_t]\mathbf{W} + \mathbf{b},
\label{eq:rnn_output} \\
\mathbf{h}^{\text{f}}_t &= \mathcal{H}(\mathbf{o}_t\,, \mathbf{h}^{\text{f}}_{t-1}), \label{eq:rnn_hidden_forward} \\
\mathbf{h}^{\text{b}}_t &= \mathcal{H}(\mathbf{o}_t\,, \mathbf{h}^{\text{b}}_{t+1}).
\label{eq:rnn_hidden_backward}
\end{align}
Here, $\mathbf{h}^{\text{f}}_t, \mathbf{h}^{\text{b}}_t \in \mathbb{R}^H$ represent the forward and backward hidden state vectors of size $H$ at recurrent time step $t$, respectively, while $\oplus$ indicates vector concatenation. 
$\mathbf{W} \in \mathbb{R}^{2H \times 2H}$ denotes a weight matrix and $\mathbf{b} \in  \mathbb{R}^{2H}$ denotes bias terms. $\mathcal{H}$ represents the hidden layer function of the recurrent layer and is realized by a Gated Recurrent Unit (GRU) \cite{Cho2014} here. We further stack multiple bidirectional GRU cells onto one another to form a deep RNN for sequential modelling as in \cite{phan2017c,Graves2013}.

\setlength\tabcolsep{2.25pt}
\begin{table}[t!]
	\caption{Configuration of the convolutional layers.}
	\footnotesize
	\begin{center}
		\begin{tabular}{|>{\arraybackslash}m{0.55in}|>{\arraybackslash}m{0.6in}|>{\centering\arraybackslash}m{0.3in}|>{\centering\arraybackslash}m{0.35in}|>{\centering\arraybackslash}m{0.25in}|>{\centering\arraybackslash}m{0.7in}|}
			\hline
			{Layer}  & {Input size} & {Filter size} & {Stride}  & {Num. filter} & {Feat. map} \parbox{0pt}{\rule{0.pt}{1ex+\baselineskip}}\\ [0ex] 	
			\hline
			Conv. 1 & $M \times T \times K$ & $5 \times 5$ & $1 \times 1$ & $64$ & $M \times T \times 64$ \parbox{0.5pt}{\rule{0pt}{1ex+\baselineskip}}\\ [0ex] 	
			Max pool. 1 & $M \times T \times 64$ & $4 \times 1$ & $4 \times 1$ & $-$ & $\frac{M}{4} \times T \times 64$ \parbox{0.5pt}{\rule{0pt}{1ex+\baselineskip}}\\ [0ex] 	
			Conv. 2 & $\frac{M}{4} \times T \times 64$ & $3 \times 3$ & $1 \times 1$ & $128$ & $\frac{M}{4} \times T \times 128$ \parbox{0.5pt}{\rule{0pt}{1ex+\baselineskip}}\\ [0ex] 	
			Max pool. 2 & $\frac{M}{4} \times T \times 128$ & $4 \times 1$ & $4 \times 1$ & $-$ & $\frac{M}{16} \times T \times 128$ \parbox{0.5pt}{\rule{0pt}{1ex+\baselineskip}}\\ [0ex] 	
			Conv. 3 & $\frac{M}{16} \times T \times 128$ & $2 \times 2$ & $1 \times 1$ & $256$ & $\frac{M}{16} \times T \times 256$ \parbox{0.5pt}{\rule{0pt}{1ex+\baselineskip}}\\ [0ex] 	
			Max pool. 3 & $\frac{M}{16} \times T \times 256$ & $4 \times 1$ & $4 \times 1$ & $-$ & $\frac{M}{64} \times T \times 256$ \parbox{0.5pt}{\rule{0pt}{1ex+\baselineskip}}\\ [0ex] 	
			
			\hline
		\end{tabular}
	\end{center}
	\label{tab:layers}
\end{table}

The recurrent output $\mathbf{Z}$ is of size $2H \times T$. In order to learn a spatio-temporal attention mask to pool and reduce $\mathbf{Z}$ into a single feature vector,
we learn two attention vectors, $\mathbf{a}^{\text{tem}} \in \mathbb{R}^T$ for temporal attention and $\mathbf{a}^{\text{spa}} \in \mathbb{R}^{2H}$ for spatial attention. Formally, the temporal attention weight $a^{\text{tem}}_t$ at the time index $t$, $1 \le t \le T$, and the spatial attention weight $a^{\text{spa}}_s$ at the spatial index $s$, $1 \le s \le 2H$  are computed as
	\begin{align}
	a^{\text{tem}}_t = \frac{\exp\big(f\left(\mathbf{z}_t\right)\big)}{\sum^{T}_{i=1}\exp\big(f(\mathbf{z}_i)\big)},
	\label{eq:temporal_attention_weight} \\
	a^{\text{spa}}_s = \frac{\exp\big(\tilde{f}\left(\mathbf{\tilde{z}}_t\right)\big)}{\sum^{2H}_{i=1}\exp\big(\tilde{f}(\mathbf{\tilde{z}}_i)\big)},
	\label{eq:spatial_attention_weight}
	\end{align}
respectively. In (\ref{eq:temporal_attention_weight}) and (\ref{eq:spatial_attention_weight}), $\mathbf{z}_t$ represents the column of $\mathbf{Z}$ at the column (i.e. temporal) index $t$ whereas $\mathbf{\tilde{z}}_s$ represents the row of $\mathbf{Z}$ at the row (i.e. spatial) index $s$. $f$ and $\tilde{f}$ denote the scoring functions of the temporal and spatial attention layers and are given by
	\begin{align}
	f(\mathbf{z}) = \tanh(\mathbf{z}^\mathsf{T}\mathbf{W}^{\text{att}} + \mathbf{b}^{\text{att}}),
	\label{eq:temproral_attention_layer} \\
	\tilde{f}(\mathbf{\tilde{z}}) = \tanh(\mathbf{\tilde{z}}^\mathsf{T}\mathbf{\tilde{W}}^{\text{att}} + \mathbf{\tilde{b}}^{\text{att}}),
	\label{eq:spatio_attention_layer}
	\end{align}
respectively, where $\mathbf{W}^{\text{att}}$ and $\mathbf{\tilde{W}}^{\text{att}}$ are the trainable weight matrices and $\mathbf{b}^{\text{att}}$ and $\mathbf{\tilde{b}}^{\text{att}}$ are the trainable biases. The spatio-temporal attention mask $\mathbf{A}$ is then obtained as
\begin{align}
	\mathbf{A} = \mathbf{a}^{\text{spa}} \otimes \mathbf{a}^{\text{tem}},
	\label{eq:spatial-temporal_mask}
\end{align}
where $\otimes$ denotes vector outer product operation.

The final feature vector $\mathbf{x}\in\mathbb{R}^{2H}$ is achieved via spatio-temporal attention pooling. Intuitively, element-wise multiplication between the recurrent output $\mathbf{Z}$ and the spatio-temporal attention mask is first carried out, followed by summation over the time dimension. Formally, the $s$-th entry, $1 \le s \le 2H$, of $\mathbf{x}$ is given as
\begin{align}
    x_s = \sum\nolimits_{t=1}^{T}\tanh(\mathbf{A}_{st}\mathbf{Z}_{st}).
	\label{eq:spatial-temporal_pooling}
\end{align}
Inspired by \cite{Zhu2018}, a $\tanh$ activation is applied prior to the summation in (\ref{eq:spatial-temporal_pooling}). Due to its output range $(-1,1)$, it is likely that $\tanh$ activation does not only suppress the irrelevant features but also enhances the informative ones in the resulting feature vector $\mathbf{x}$ \cite{Zhu2018}. 

Eventually, the obtained feature vector $\mathbf{x}$ is presented to a softmax layer to accomplish classification. 

\subsection{Calibration with Support Vector Machine}
Compared to the standard softmax, Support Vector Machines (SVM) usually achieve better generalization due to their maximum margin property \cite{Boser1992}. Similar to \cite{phan2017b,phan2017c}, after training the network, we calibrate the final classifier by employing a linear SVM in replacement for the softmax layer. The trained network is used to extract feature vectors for the original training examples (without data augmentation) which are used to train the SVM classifier. During testing, the SVM classifier is subsequently used to classify those feature vectors extracted for the test examples. The raw SVM scores are also calibrated and converted into a proper
posterior probability as in \cite{Platt1999}.

\section{Between-class data augmentation and KL-divergence loss}

In deep learning, data augmentation, which is to increase the data variation by altering the property of the genuine data, is an important method to improve performance of the task at hand. Techniques like adding background noise \cite{Tokozume2017, Salamon2017}, pitch shifting \cite{Salamon2017}, and sample mixing \cite{Tokozume2018,Xu2018}, have been proven to be useful for environmental sound recognition in general. Motivated by the work of Tokozume \emph{et al.} \cite{Tokozume2018}, we pursue a \emph{between-class} (BC) data augmentation approach that mixes two samples of different classes with a random factor to generate BC examples for network training.

Let $\mathbf{S}_1$ and $\mathbf{S}_2$ denote two samples of two different classes and let $\mathbf{y}_1$ and $\mathbf{y}_2$ denote their corresponding one-hot labels. A random factor $r \thicksim U(0,1)$ is then generated and used to mix the two samples and their labels to create a new BC sample $\mathbf{S}^\text{BC}$ and its labels $\mathbf{y}^\text{BC}$:
\begin{align}
    \mathbf{S}^\text{BC} = \frac{r\mathbf{S}_{1} + (1-r)\mathbf{S}_{2}}{\sqrt{r^2 + (1-r)^2}}, \label{eq:bc_sample} \\
    \mathbf{y}^\text{BC} = r\mathbf{y}_{1} + (1-r)\mathbf{y}_{2}.
	\label{eq:bc_label}
\end{align}
Note that the BC label $\mathbf{y}^\text{BC}$ is no longer a one-hot label but still a proper probability distribution, which represents the amplitude of the constituents $\mathbf{S}_1$ and $\mathbf{S}_2$ in the between-class sample $\mathbf{S}^\text{BC}$. For example, mixing a \emph{restaurant} scene and a \emph{tubestation} scene with a factor $r = 0.3$ will result in a label \{\emph{restaurant}: 0.3, \emph{tubestation}: 0.7\}. Training a network with the BC sample $(\mathbf{S}^\text{BC}, \mathbf{y}^\text{BC})$, we expect the network's class probability distribution output $\mathbf{\hat{y}}$ to be as close to the $\mathbf{y}^\text{BC}$ as possible. Therefore, KL-divergence between $\mathbf{y}^\text{BC}$ and $\hat{y}$ is used as the network loss:
\begin{align}
    L = D_{\text{KL}}(\mathbf{y}^\text{BC} || \mathbf{\hat{y}}) = \sum\nolimits_{c=1}^C y^{\text{BC}}_i \log\frac{y^{\text{BC}}_c}{\hat{y}_c},
	\label{eq:kl_loss}
\end{align}
where $C$ is the number of classes. Learning with between-class examples was shown to enlarge Fisher's criterion, i.e. the ratio of the between-class distance to the within-class variance \cite{Tokozume2018}.

\section{Experiments}

\subsection{LITIS-Rouen dataset}
\label{ssec:dataset}
The LITIS-Rouen dataset consists of 3026 examples of 19 scene categories \cite{Rakotomamonjy2015}. Each class is specific to a location such as a train station or an open market. The audio recordings have a duration of 30 seconds and a sampling rate of 22050 Hz. The dataset has a total duration of 1500 minutes. 
We follow the training/testing splits in the seminal work \cite{Rakotomamonjy2015} and report average performances over 20 splits.

\subsection{Features}
\label{ssec:features}
A 2-second audio segment, sampled at $fs = 22050$ Hz, was transformed into a log Mel-scale spectrogram with $M=64$ Mel-scale filters in the frequency range from 50 Hz to Nyquist rate. A frame size of 50 ms with 50\% overlap was used, resulting in $T = 80$ frames in total. Likewise, another image was produced for the auxiliary channel (cf. Section \ref{ssec:input}). All in all, we obtained a multi-channel image $\mathbf{S}\in \mathbb{R}^{M\times T \times K}$ where $K=2$ denotes the number of channels.

Beside log Mel-scale spectrogram, we also studied log Gammatone spectrogram in this work. Repeating a similar feature extraction procedure using $M=64$ Gammatone filters in replacement of the above-mentioned Mel-scale filters, we obtained a multi-channel log Gammatone spectrogram image for a 2-second audio segment.

\subsection{Network parameters}
The studied networks were implemented using \emph{Tensorflow} framework \cite{Abadi2016}. We applied \emph{dropout} \cite{Srivastava2014} to the convolutional layers described in Section \ref{ssec:convolution} with a dropout rate of $0.25$. The GRU cells used to realize two bidirectional recurrent layers in Section \ref{ssec:recurrent} have their hidden size $H=128$, and a dropout rate of $0.1$ was commonly applied to their inputs and outputs. Both the spatial and temporal attention layers have the same size of 64. The networks were trained for 500 epochs with a minibatch size of 100. \emph{Adam} optimizer \cite{Kingma2015} was used for network training with a learning rate of $10^{-4}$.

Finally, the trade-off parameter $C$ for the SVM classifier used for calibration was fixed at 0.1.
\subsection{Baseline}

In order to illustrate the efficiency of the recurrent layers with spatio-temporal attention pooling, we used the CNN block in Figure \ref{fig:overview} as a deep CNN baseline. For this baseline, global max pooling was used after the last pooling layer to derive the final feature vector for classification. Other configuration settings were the same as for the proposed CRNN with spatio-temporal attention pooling.

\subsection{Experimental results}

Table \ref{tab:comparison_baseline} shows the classification accuracy obtained by the proposed network (referred to as Att-CRNN) and the CNN baseline. Note that the classification label of a 30-second recording was derived via aggregation of the classification results of its 2-second segments. To this end, probabilistic multiplicative fusion, followed by likelihood maximization were carried out similar to \cite{phan2017c}. 

\begin{figure} [!t]
	\centering
	\includegraphics[width=1\linewidth]{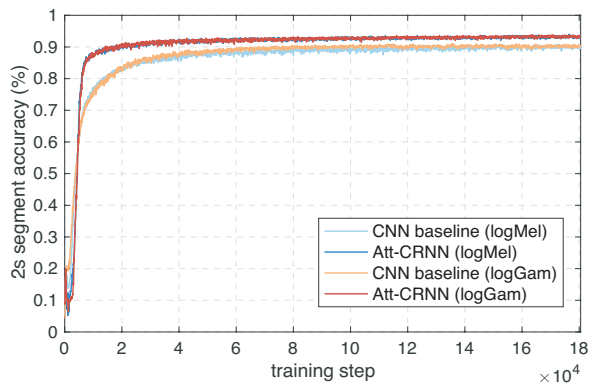}
	\caption{Variation of the 2-second segment test accuracy during network training. The first cross-validation fold is shown as representative here.}
	\label{fig:learning_curve}
\end{figure}

Overall, the proposed Att-CRNN outperforms the CNN baseline regardless of the features used, improving the accuracy on 2-second segment classification by $1.45\%$ and $1.52\%$ absolute using log Mel-scale and log Gammatone spectrograms, respectively. The better generalization of the proposed Att-CRNN over the CNN baseline can also be seen via patterns of their test accuracy curves during network training as shown in Figure \ref{fig:learning_curve} for the first cross-validation fold. In turn, the improvements on the 2-second segment classification led to $0.42\%$ and $0.59\%$ absolute gains on the 30-second recordings classification using log Mel-scale and log Gammatone spectrograms, respectively. These are equivalent to a relative classification error reduction of $22.22\%$ and $27.70\%$, respectively. 

\setlength\tabcolsep{2.25pt}
\begin{table}[t!]
	\caption{Classification accuracy obtained by the proposed Att-CRNN and the CNN baseline.}
	\begin{center}
		\begin{tabular}{|>{\arraybackslash}m{1.4in}|>{\centering\arraybackslash}m{0.5in}|>{\centering\arraybackslash}m{0.5in}|}
			\hline
			{\bf System}  & {\bf 2s} & {\bf 30s}  \parbox{0pt}{\rule{0pt}{1ex+\baselineskip}}\\ [0ex] 	
			\hline
			Att-CRNN (logMel)  & $93.78$ & $98.53$ \parbox{0pt}{\rule{0pt}{0.5ex+\baselineskip}}\\ [0ex] 	
			Att-CRNN (logGam) & $93.65$ & $98.46$  \parbox{0pt}{\rule{0pt}{0.5ex+\baselineskip}}\\ [0ex] 	
			CNN baseline (logMel) & $92.33$ & $98.11$  \parbox{0pt}{\rule{0pt}{0.5ex+\baselineskip}}\\ [0ex] 	
			CNN baseline (logGam) & $92.13$ & $97.87$  \parbox{0pt}{\rule{0pt}{0.5ex+\baselineskip}}\\ [0ex] 	
			\hline
		\end{tabular}
	\end{center}
	\label{tab:comparison_baseline}
\end{table}

\subsection{Performance comparison with state-of-the-art}

The experimental LITIS Rouen dataset has been extensively evaluated in literature. Table \ref{tab:comparison_litis} provides a comprehensive comparison between the performance obtained by the proposed Att-CRNN and the CNN baseline to those reported in previous works in terms of overall accuracy, average F1-score, and average precision. Overall, this comparison shows that our presented systems obtain better performance than all other counterparts. 
On the one hand, despite being simple, the CNN baseline alone performs comparably well compared to the state-of-the-art system, i.e. Temporal Transformer CNN \cite{Zhang2018}, likely due to the positive effect of the between-class data augmentation. On the other hand, the proposed Att-CRNN with log Mel-scale and log Gammatone spectrogram as features improves the accuracy by $0.47\%$ and $0.40\%$ over the state-of-the-art system, respectively. These accuracy gains are equivalent to a relative classification error reduction of $24.2\%$ and $20.6\%$, respectively. Combining the classification results of the Att-CRNN on both types of feature using the probabilistic multiplicative aggregation \cite{phan2017c} (i.e. Att-CRNN (fusion) in Table \ref{tab:comparison_litis}) further enlarges the margin up to $0.66\%$ absolute gain on the overall accuracy, or $34.02\%$ on relative classification error reduction.

\section{Conclusions}
This paper has presented an approach for audio scene classification using spatio-temporal attention pooling in combination with convolutional recurrent neural networks. The convolutional layers in this network are expected to learn invariant features from the input whose temporal dynamics are further encoded by bidirectional recurrent layers. Attention layers then learn attention weight vectors in the spatial and temporal dimensions from the recurrent output, which collectively construct a spatio-temporal attention mask able to weigh and pool the recurrent output into a single feature vector for classification. The proposed network was trained with between-class examples and KL-divergence loss. Evaluated on the LITIS Rouen dataset, the proposed method achieved good classification performance, outperforming a strong CNN baseline as well as the previously published state-of-the-art systems.

\setlength\tabcolsep{2.25pt}
\begin{table}[t!]
	\caption{Performance comparison on the LITIS Rouen dataset. We mark in bold where the performances achieved by our proposed systems are better than all those of previous works.}
	\begin{center}
		\begin{tabular}{|>{\arraybackslash}m{1.6in}|>{\centering\arraybackslash}m{0.35in}|>{\centering\arraybackslash}m{0.5in}|>{\centering\arraybackslash}m{0.4in}|}
			\hline
			{\bf System}  & {\bf Acc.} & {\bf F1-score}  & {\bf Prec.} \parbox{0pt}{\rule{0pt}{1ex+\baselineskip}}\\ [0ex] 	
			\hline
			\emph{Att-CRNN (fusion)}  & $\bm{98.72}$ & $\bm{98.57}$ & $\bm{98.40}$   \parbox{0pt}{\rule{0pt}{0.5ex+\baselineskip}}\\ [0ex] 	
			\emph{Att-CRNN (logMel)}  & $\bm{98.53}$ & $\bm{98.39}$ & $\bm{98.20}$   \parbox{0pt}{\rule{0pt}{0.5ex+\baselineskip}}\\ [0ex] 	
			\emph{Att-CRNN (logGam)} & $\bm{98.46}$ & $\bm{98.28}$ & $\bm{98.10}$   \parbox{0pt}{\rule{0pt}{0.5ex+\baselineskip}}\\ [0ex] 	
			\emph{CNN baseline (fusion)} & $\bm{98.17}$ & $\bm{97.92}$ & $\bm{97.71}$   \parbox{0pt}{\rule{0pt}{0.5ex+\baselineskip}}\\ [0ex] 	
			\emph{CNN baseline (logMel)} & $\bm{98.11}$ & $\bm{97.82}$ & $\bm{97.63}$   \parbox{0pt}{\rule{0pt}{0.5ex+\baselineskip}}\\ [0ex] 	
			\emph{CNN baseline (logGam)} & $97.87$ & $97.59$ & $97.39$   \parbox{0pt}{\rule{0pt}{0.5ex+\baselineskip}}\\ [0ex] 	
			\hline
			\hline
			Temp. Transformer CNN \cite{Zhang2018} & $98.06$ & $-$ & $-$   \parbox{0pt}{\rule{0pt}{0.5ex+\baselineskip}}\\ [0ex] 	
			Temp. Transformer LSTM \cite{Zhang2018} & $97.86$ & $-$ & $-$ \parbox{0pt}{\rule{0pt}{0.5ex+\baselineskip}}\\ [0ex] 	
			LSTM-SA \cite{Zhang2018b} & $97.92$ & $-$ & $-$ \parbox{0pt}{\rule{0pt}{0.5ex+\baselineskip}}\\ [0ex] 	
			LTE-RNN \cite{phan2017c} & $97.8$  & $97.7$ & $97.5$ \parbox{0pt}{\rule{0pt}{0.5ex+\baselineskip}}\\ [0ex] 	
			Temp. Transformer DNN \cite{Zhang2018} & $97.40$ & $-$ & $-$  \parbox{0pt}{\rule{0pt}{0.5ex+\baselineskip}}\\ [0ex] 	
			
			CQT+HOG \cite{Ye2018} & $97.0$ & $-$ & $-$  \parbox{0pt}{\rule{0pt}{0.5ex+\baselineskip}}\\ [0ex] 	
			
			LTE-CNN \cite{phan2017b}  & $96.6$  & $96.5$ & $96.3$ \parbox{0pt}{\rule{0pt}{0.5ex+\baselineskip}}\\ [0ex] 	%
			Scene-LTE + Speech-LTE \cite{phan2016d} & $96.4$  & $96.2$ & $95.9$  \parbox{0pt}{\rule{0pt}{0.5ex+\baselineskip}}\\ [0ex] 	
			
			1D+2D+3D CNNs \cite{Yin2018} & $96.4$  & $-$ & $-$  \parbox{0pt}{\rule{0pt}{0.5ex+\baselineskip}}\\ [0ex] 	
			
			{FisherHOG+ProbSVM \cite{Ye2015}}   & $96.0$ & $-$ & $-$ \parbox{0pt}{\rule{0pt}{0.5ex+\baselineskip}}\\ [0ex]
			{Kernel PCA \cite{Bisot2016}}  & $-$ & $95.6$ & $-$ \parbox{0pt}{\rule{0pt}{0.5ex+\baselineskip}}\\ [0ex]
			{Convolutive NMF \cite{Bisot2016}}  & $-$ & $94.5$ & $-$ \parbox{0pt}{\rule{0pt}{0.5ex+\baselineskip}}\\ [0ex]
			{Sparse NMF \cite{Bisot2016}}  & $-$ & $94.1$ & $-$ \parbox{0pt}{\rule{0pt}{0.5ex+\baselineskip}}\\ [0ex]
			{HOG+SPD \cite{Bisot2015}}   & $93.4$ & $92.8$  & $93.3$
			 \parbox{0pt}{\rule{0pt}{0.5ex+\baselineskip}}\\ [0ex]
			 {MFCC+DNN \cite{Petetin2015}} & $-$ & $-$ & $92.2$  \parbox{0pt}{\rule{0pt}{0.5ex+\baselineskip}}\\ [0ex]
			{HOG \cite{Rakotomamonjy2015}}  & $-$ & $-$ & $91.7$ \parbox{0pt}{\rule{0pt}{0.5ex+\baselineskip}}\\ [0ex]
			\hline
		\end{tabular}
	\end{center}
	\label{tab:comparison_litis}
\end{table}

\section{Acknowledgements}

We gratefully acknowledge the support of NVIDIA Corporation with the donation of the Titan V GPU used for this research and Specialist and High Performance Computing systems provided by Information Services at the University of Kent.

\newpage
\bibliographystyle{IEEEtran}
\bibliography{reference}

\begin{thebibliography}{10}
\providecommand{\url}[1]{#1}
\csname url@samestyle\endcsname
\providecommand{\newblock}{\relax}
\providecommand{\bibinfo}[2]{#2}
\providecommand{\BIBentrySTDinterwordspacing}{\spaceskip=0pt\relax}
\providecommand{\BIBentryALTinterwordstretchfactor}{4}
\providecommand{\BIBentryALTinterwordspacing}{\spaceskip=\fontdimen2\font plus
\BIBentryALTinterwordstretchfactor\fontdimen3\font minus
  \fontdimen4\font\relax}
\providecommand{\BIBforeignlanguage}[2]{{%
\expandafter\ifx\csname l@#1\endcsname\relax
\typeout{** WARNING: IEEEtran.bst: No hyphenation pattern has been}%
\typeout{** loaded for the language `#1'. Using the pattern for}%
\typeout{** the default language instead.}%
\else
\language=\csname l@#1\endcsname
\fi
#2}}
\providecommand{\BIBdecl}{\relax}
\BIBdecl

\bibitem{Mesaros2018}
A.~Mesaros, T.~Heittola, E.~Benetos, P.~Foster, M.~Lagrange, T.~Virtanen, and
  M.~D. Plumbley, ``Detection and classification of acoustic scenes and events:
  Outcome of the {DCASE} 2016 challenge,'' \emph{IEEE/ACM Transactions on
  Audio, Speech, and Language Processing (TASLP)}, vol.~26, no.~2, pp.
  379--393, 2018.

\bibitem{phan2017c}
H.~Phan, P.~Koch, F.~Katzberg, M.~Maass, R.~Mazur, and A.~Mertins, ``Audio
  scene classification with deep recurrent neural networks,'' in \emph{Proc.
  Interspeech}, 2017, pp. 3043--3047.

\bibitem{Mun2017a}
S.~Mun, S.~Park, D.~Han, and H.~Ko, ``Generative adversarial network based
  acoustic scene training set augmentation and selection using {SVM}
  hyper-plane,'' in \emph{Proc. DCASE Workshop}, 2017.

\bibitem{Han2017}
Y.~Han, J.~Park, and K.~Lee, ``Convolutional neural networks with binaural
  representations and background subtraction for acoustic scene
  classification,'' in \emph{Proc. DCASE Workshop}, 2017.

\bibitem{Han2016}
Y.~Han and K.~Lee, ``Convolutional neural network with multiple-width
  frequency-delta data augmentation for acoustic scene classification,''
  DCASE2016 Challenge, Tech. Rep., September 2016.

\bibitem{phan2017b}
H.~Phan, L.~Hertel, M.~Maass, P.~Koch, R.~Mazur, and A.~Mertins, ``Improved
  audio scene classification based on label-tree embeddings and convolutional
  neural networks,'' \emph{IEEE/ACM Trans. on Audio, Speech and Language
  Processing (TASLP)}, vol.~25, no.~6, pp. 1278--1290, 2017.

\bibitem{Valenti2016}
M.~Valenti, A.~Diment, G.~Parascandolo, S.~Squartini, and T.~Virtanen, ``{DCASE
  2016} acoustic scene classification using convolutional neural networks,''
  Detection and Classification of Acoustic Scenes and Events 2016, Tech. Rep.,
  2016.

\bibitem{phan2017a}
H.~Phan, P.~Koch, L.~Hertel, M.~Maass, R.~Mazur, and A.~Mertins, ``{CNN-LTE:} a
  class of {1-X} pooling convolutional neural networks on label tree embeddings
  for audio scene classification,'' in \emph{Proc. ICASSP}, 2017.

\bibitem{Yin2018}
Y.~Yin, R.~R. Shah, and R.~Zimmermann, ``Learning and fusing multimodal deep
  features for acoustic scene categorization,'' in \emph{Proc. ACMMM}, 2018,
  pp. 1892--1900.

\bibitem{Zhang2018}
T.~Zhang, K.~Zhang, and J.~Wu, ``Temporal transformer networks for acoustic
  scene classification,'' in \emph{Proc. Interspeech}, 2018, pp. 1349--1353.

\bibitem{Nguyen2018}
T.~Nguyen and F.~Pernkopf, ``Acoustic scene classification using a
  convolutional neural network ensemble and nearest neighbor filters,'' in
  \emph{Proc. DCASE Workshop}, 2018.

\bibitem{Vu2016}
T.~H. Vu and J.-C. Wang, ``Acoustic scene and event recognition using recurrent
  neural networks,'' Detection and Classification of Acoustic Scenes and Events
  2016, Tech. Rep., 2016.

\bibitem{Guo2017}
J.~Guo, N.~Xu, L.-J. Li, and A.~Alwan, ``Attention based {CLDNNs} for
  short-duration acoustic scene classification,'' in \emph{Proc. Interspeech},
  2017.

\bibitem{Phan2019c}
H.~Phan, O.~Y. Ch\'{e}n, P.~Koch, L.~Pham, I.~McLoughlin, A.~Mertins, and
  M.~{De Vos}, ``Beyond equal-length snippets: How long is sufficient to
  recognize an audio scene?'' in \emph{Proc. 2019 AES Conference on Audio
  Forensics}, 2019.

\bibitem{Luong2015b}
T.~Luong, H.~Pham, and C.~D. Manning, ``Effective approaches to attention-based
  neural machine translation,'' in \emph{Proc. EMNLP}, 2015, pp. 1412--1421.

\bibitem{Cakir2017}
E.~\c{C}akir, G.~Parascandolo, T.~Heittola, H.~Huttunen, and T.~Virtanen,
  ``Convolutional recurrent neural networks for polyphonic sound event
  detection,'' \emph{IEEE/ACM Trans. on Audio, Speech, and Language
  Processing}, vol.~5, no.~6, pp. 1291--1303, 2017.

\bibitem{Phan2019b}
H.~Phan, O.~Y. Ch\'{e}n, P.~Koch, L.~Pham, I.~McLoughlin, A.~Mertins, and M.~D.
  Vos, ``Unifying isolated and overlapping audio event detection with
  multi-label multi-task convolutional recurrent neural networks,'' in
  \emph{Proc. ICASSP}, 2019.

\bibitem{Tokozume2018}
Y.~Tokozume, Y.~Ushiku, and T.~Harada, ``Learning from between-class examples
  for deep sound recognition,'' in \emph{Proc. ICLR}, 2018.

\bibitem{Rakotomamonjy2015}
A.~Rakotomamonjy and G.~Gasso, ``Histogram of gradients of time-frequency
  representations for audio scene classification,'' \emph{IEEE/ACM Trans.
  Audio, Speech, and Language Processing}, vol.~23, no.~1, pp. 142--153, 2015.

\bibitem{Martin2001}
R.~Martin, ``Noise power spectral density estimation based on optimal smoothing
  and minimum statistics,'' \emph{IEEE Trans. on Speech and Audio Processing},
  vol.~9, no.~5, pp. 504--512, 2001.

\bibitem{Nair2010}
V.~Nair and G.~E. Hinton, ``Rectified linear units improve restricted
  {Boltzmann} machines,'' in \emph{Proc. ICML}, 2010.

\bibitem{Ioffe2015}
S.~Ioffe and C.~Szegedy, ``Batch normalization: Accelerating deep network
  training by reducing internal covariate shift,'' in \emph{Proc. ICML}, 2015,
  pp. 448--456.

\bibitem{Cho2014}
K.~Cho, B.~{van Merrienboer}, C.~Gulcehre, F.~Bougares, H.~Schwenk, and
  Y.~Bengio, ``Learning phrase representations using {RNN} encoder-decoder for
  statistical machine translation,'' in \emph{Proc. EMNLP}, 2014, pp.
  1724--1734.

\bibitem{Graves2013}
A.~Graves, A.~Mohamed, and G.~Hinton, ``Speech recognition with deep recurrent
  neural networks,'' in \emph{Proc. ICASSP}, 2013, pp. 6645--6649.

\bibitem{Zhu2018}
C.~Zhu, X.~Tan, F.~Zhou, X.~Liu, K.~Yue, E.~Ding, and Y.~Ma, ``Fine-grained
  video categorization with redundancy reduction attention,'' in \emph{Proc.
  ECCV}, 2018, pp. 139--155.

\bibitem{Boser1992}
B.~E. Boser, I.~M. Guyon, and V.~N. Vapnik, ``A training algorithm for optimal
  margin classifiers,'' in \emph{Proc. COLT}, 1992, pp. 144--152.

\bibitem{Platt1999}
J.~Platt, \emph{Advances in Large Margin Classifiers}.\hskip 1em plus 0.5em
  minus 0.4em\relax MIT Press, 1999, ch. Probabilistic outputs for support
  vector machines and comparisons to regularized likelihood methods.

\bibitem{Tokozume2017}
Y.~Tokozume and T.~Harada, ``Learning environmental sounds with end-to-end
  convolutional neural network,'' in \emph{Proc. ICASSP}, 2017, pp. 2721--2725.

\bibitem{Salamon2017}
J.~Salamon and J.~P. Bello, ``Deep convolutional neural networks and data
  augmentation for environmental sound classification,'' \emph{IEEE Signal
  Processing Letters}, vol.~24, no.~3, pp. 279--283, 2017.

\bibitem{Xu2018}
K.~Xu, D.~Feng, H.~Mi, B.~Zhu, D.~Wang, L.~Zhang, H.~Cai, and S.~Liu,
  ``Mixup-based acoustic scene classification using multi-channel convolutional
  neural network,'' \emph{arXiv preprint arXiv:1805.07319}, 2018.

\bibitem{Abadi2016}
M.~{Abadi \emph{et al.}}, ``Tensorflow: Large-scale machine learning on
  heterogeneous distributed systems,'' \emph{arXiv:1603.04467}, 2016.

\bibitem{Srivastava2014}
N.~Srivastava, G.~Hinton, A.~Krizhevsky, I.~Sutskever, and R.~Salakhutdinov,
  ``Dropout: A simple way to prevent neural networks from overfitting,''
  \emph{Journal of Machine Learning Research (JMLR)}, vol.~15, pp. 1929--1958,
  2014.

\bibitem{Kingma2015}
D.~P. Kingma and J.~L. Ba, ``Adam: a method for stochastic optimization,'' in
  \emph{Proc. ICLR}, 2015, pp. 1--13.

\bibitem{Zhang2018b}
T.~Zhang, K.~Zhang, and J.~Wu, ``Data independent sequence augmentation method
  for acoustic scene classification,'' in \emph{Proc. Interspeech}, 2018.

\bibitem{Ye2018}
J.~Ye, T.~Kobayashi, N.~Toyama, H.~Tsuda, and M.~Murakawa, ``Acoustic scene
  classification using efficient summary statistics and multiple
  spectro-temporal descriptor fusion,'' \emph{Applied Sciences}, vol.~8, p.
  1363, 2018.

\bibitem{phan2016d}
H.~Phan, L.~Hertel, M.~Maass, P.~Koch, and A.~Mertins, ``Label tree embeddings
  for acoustic scene classification,'' in \emph{Proc. ACMMM}, 2016, pp.
  486--490.

\bibitem{Ye2015}
J.~Ye, T.~Kobayashi, M.~Murakawa, and T.~Higuchi, ``Acoustic scene
  classification based on sound textures and events,'' in \emph{Proc. ACM
  Multimedia}, 2015, pp. 1291--1294.

\bibitem{Bisot2016}
V.~Bisot, R.~Serizel, S.~Essid, and G.~Richard, ``Acoustic scene classification
  with matrix factorization for unsupervised feature learning,'' in \emph{Proc.
  ICASSP}, 2016, pp. 6445--6449.

\bibitem{Bisot2015}
V.~Bisot, S.~Essid, and G.~Richard, ``{HOG} and subband power distribution
  image features for acoustic scene classification,'' in \emph{Proc. EUSIPCO},
  2015, pp. 719--723.

\bibitem{Petetin2015}
Y.~Petetin, C.~Laroche, and A.~Mayoue, ``Deep neural networks for audio scene
  recognition,'' in \emph{Proc. EUSIPCO}, 2015, pp. 125--129.

\end{thebibliography}

\end{document}